\documentclass{IEEEtran}
\usepackage{amsmath,graphicx}
\usepackage{amssymb}
\usepackage{amsthm}
\usepackage{algorithm}
\usepackage{algorithmic}
\usepackage{color}
\usepackage{array}
\usepackage{subfigure}
\usepackage{url}
\usepackage{soul}
\usepackage[flushleft]{threeparttable} 

\title{Spectrum Sharing in Vehicular Networks Based on Multi-Agent Reinforcement Learning}
\author{
Le~Liang,~\IEEEmembership{Member,~IEEE}, Hao Ye,~\IEEEmembership{Student Member,~IEEE}, and Geoffrey Ye Li,~\IEEEmembership{Fellow,~IEEE}
\thanks{
This work was supported in part by a research gift from Intel Corporation and in part by the National Science Foundation under Grants 1731017 and 1815637.
}
\thanks{L. Liang is with Intel Labs, Hillsboro, OR 97124. The work was done when he was with the School of Electrical and Computer Engineering, Georgia Institute of Technology, Atlanta, GA 30339 USA (e-mail: lliang@gatech.edu).}
\thanks{
H. Ye and G. Y. Li are with the School of Electrical and Computer Engineering, Georgia Institute of Technology, Atlanta, GA 30339 USA (e-mail: yehao@gatech.edu; liye@ece.gatech.edu).
}
}

\begin{document}

\maketitle
\begin{abstract}
This paper investigates the spectrum sharing problem in vehicular networks based on multi-agent reinforcement learning, where multiple vehicle-to-vehicle (V2V) links reuse the frequency spectrum preoccupied by vehicle-to-infrastructure (V2I) links.
Fast channel variations in high mobility vehicular environments preclude the possibility of collecting accurate instantaneous channel state information at the base station for centralized resource management.
In response, we model the resource sharing as a multi-agent reinforcement learning problem, which is then solved using a fingerprint-based deep Q-network method that is amenable to a distributed implementation.
The V2V links, each acting as an agent, collectively interact with the communication environment, receive distinctive observations yet a common reward, and learn to improve spectrum and power allocation through updating Q-networks using the gained experiences.
We demonstrate that with a proper reward design and training mechanism, the multiple V2V agents successfully learn to cooperate in a distributed way to simultaneously improve the sum capacity of V2I links and payload delivery rate of V2V links.

\end{abstract}

\begin{IEEEkeywords}
Vehicular networks, distributed spectrum access, spectrum and power allocation, multi-agent reinforcement learning.
\end{IEEEkeywords}

\section{Introduction}\label{marl:sec:intro}
Vehicular communication, commonly referred to as vehicle-to-everything (V2X) communication, is envisioned to transform connected vehicles and intelligent transportation services in various aspects, such as road safety, traffic efficiency, and ubiquitous Internet access \cite{Liang2017vehicular,Peng2018vehicular}.
More recently, the 3rd Generation Partnership Project (3GPP) has been looking to support V2X services in long-term evolution (LTE) and future 5G cellular networks \cite{3GPPr14v2x,3GPPr15v2x,Molina2017LTEV}.
Cross-industry consortium, such as the 5G automotive association (5GAA), has been founded by telecommunication and automotive industries to push development, testing, and deployment of cellular V2X technologies.

\subsection{\textcolor{black}{{Problem Statement and Motivation}}}

\begin{figure}[!t]
\centering
\includegraphics[width=0.99\linewidth]{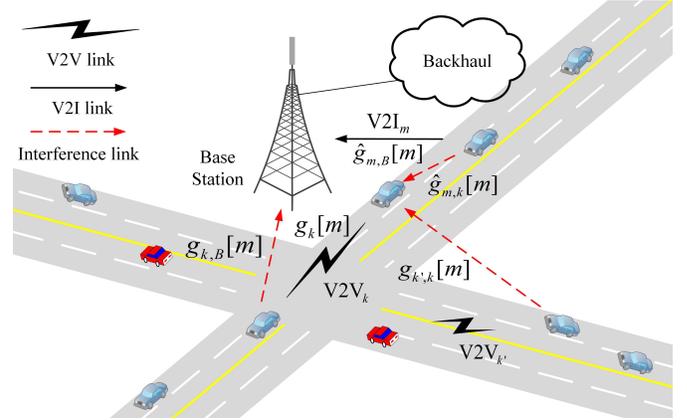}
\caption{An illustrative structure of vehicular networks, where V2I and V2V links are indexed by $m$ and $k$ (or $k'$), respectively. Each of the V2I links is preassigned an orthogonal spectrum sub-band and hence the sub-band is also indexed by $m$. }\label{marl:fig:v2x}
\end{figure}



\textcolor{black}{{This paper considers spectrum access design in vehicular networks, which in general comprise both vehicle-to-infrastructure (V2I) and vehicle-to-vehicle (V2V) connectivity.
As illustrated in Fig.~\mbox{\ref{marl:fig:v2x}}, the V2I links connect each vehicle to the base station (BS) or BS-type road side unit (RSU) while V2V links provide direct communications among neighboring vehicles.
We focus on the cellular based V2X architecture discussed within the 3GPP \mbox{\cite{3GPPr14v2x}}, where V2I and V2V connections are supported through cellular (Uu) and sidelink (PC5) radio interfaces, respectively.
A wide array of new use cases and requirements have been proposed and analyzed for 5G V2X enhancements in Release 15 \mbox{\cite{3GPPr15v2x,Molina2017LTEV}}.
For example, the 5G cellular V2X networks are required to provide simultaneous support for mobile high data rate entertainment and advanced driving in 5G cellular V2X networks.
The entertainment applications require high bandwidth V2I connection to the BS (and further the Internet) for, e.g., video streaming.
Meanwhile, the advanced driving service needs to periodically disseminate safety messages among neighboring vehicles (e.g., 10, 20, 50 packets per second depending on vehicle mobility \mbox{\cite{Molina2017LTEV}})
through V2V communications, with high reliability.
The safety messages usually include such information as vehicle position, speed, heading, etc. to increase ``co-operative awareness'' of the local driving environment for all vehicles.
}}


\textcolor{black}{{This work is based on Mode 4 defined in the 3GPP cellular V2X architecture, where the vehicles have a pool of radio resources that they can autonomously select from for V2V communications \mbox{\cite{Molina2017LTEV}}.
To fully use available resources, we propose that such sidelink V2V connections share spectrum with Uu (V2I) links with necessary interference management design.
We make some simplification on the V2I communications in that they have preoccupied the spectrum in an orthogonal way with fixed transmission power.
Hence, the resource optimization is left for the design of V2V connections that need to devise effective strategies of spectrum sharing with V2I links, including the selection of spectrum sub-band and proper control of transmission power, to meet the diverse service requirements of both V2I and V2V links.
Such an architecture provides more opportunities for the coexistence of V2I and V2V connections on limited frequency spectrum, but also complicates interference design in the network and hence motivates this work.
}}

\textcolor{black}{{While there exists a rich body of literature applying conventional optimization methods to solve similarly formulated V2X resource allocation problems, they actually find difficulty to fully address them in several aspects.
On one hand, fast changing channel conditions in vehicular environments causes substantial uncertainty for resource allocation, e.g., in terms of performance loss induced by inaccuracy of acquired channel state information (CSI).
On the other hand, increasingly diverse service requirements are being brought up to support new V2X applications, such as simultaneously maximizing throughput and reliability for a mix of V2X traffic, as discussed earlier in the motivational example.
Such requirements are sometimes hard to be modeled in a mathematically exact way, not to mention a systematic approach to find optimal solutions.
Fortunately, reinforcement learning (RL) has been shown effective in addressing decision making under uncertainty \mbox{\cite{sutton1998reinforcement}}. In particular, recent success of deep RL in human-level video game play \mbox{\cite{mnih2015human} and AlphaGo \cite{silver2016mastering}} has sparked a flurry of interest in applying RL techniques to solve problems from a wide variety of areas and remarkable progress has been made ever since \mbox{\cite{He2017deep,He2018secure,Mao2016resource}}.
It provides a robust and principled way to treat environment dynamics and perform sequential decision making under uncertainty, thus representing a promising method to handle the unique and challenging V2X dynamics.
In addition, the hard-to-optimize objective issues can also be nicely addressed in a RL framework through designing training rewards such that they correlate with the final objective. The learning algorithm can then figure out a clever strategy to approach the ultimate goal by itself.
Another potential advantage of using RL for resource allocation is that distributed algorithms are made possible, as demonstrated in \mbox{\cite{Ye2019deep}}, which treats each V2V link as an agent that learns to refine its resource sharing strategy through interacting with the unknown vehicular environment.
As a result, we investigate the use of multi-agent RL tools to solve the V2X spectrum access problem in this work.
}}

\subsection{\textcolor{black}{{Related Work}}}
To address the challenges caused by fleeting channel conditions in vehicular environments, a heuristic spatial spectrum reuse scheme has been developed in \cite{Botsov2014location} for device-to-device (D2D) based vehicular networks, relieving requirements on full CSI.
In \cite{Sun2016radio}, V2X resource allocation, which maximizes throughput of V2I links, adapts to slowly-varying large-scale channel fading and hence reduces network signaling overhead. Further in \cite{Sun2016cluster}, similar strategies have been adopted while spectrum sharing is allowed not only between V2I and V2V links but also among peer V2V links.
A proximity and QoS-aware resource allocation scheme for V2V communications has been developed in \cite{Ashraf2018dynamic} that minimizes the total transmission power of all V2V links while satisfying latency and reliability requirements using a Lyapunov-based stochastic optimization framework.
Sum ergodic capacity of V2I links has been maximized with V2V reliability guarantee using large-scale fading channel information in \cite{Liang2017resource} or CSI from periodic feedback in \cite{Liang2017spectrum}.
A novel graph-based approach has been further developed in \cite{Liang2018graph} to deal with a generic V2X resource allocation problem.

\textcolor{black}{{Apart from the traditional optimization methods, RL based approaches have been developed in several recent works to address resource allocation in V2X networks \mbox{\cite{Ye2018mlv2x, Liang2018toward}}.}}
In \cite{salahuddin2016reinforcement}, RL algorithms have been applied to address the resource provisioning problem in vehicular clouds such that dynamic resource demands and stringent quality of service requirements of various entities in the clouds are met with minimal overhead.
The radio resource management problem for transmission delay minimization in software-defined vehicular networks has been studied in \cite{zheng2016delay}, which is formulated as an infinite-horizon partially observed Markov decision process (MDP) and solved with an online distributed learning algorithm based on an equivalent Bellman equation and stochastic approximation.
In \cite{He2017integrated}, a deep RL based method has been proposed to jointly manage the networking, caching, and computing resources in virtualized vehicular networks with information-centric networking and mobile edge computing capabilities. The developed deep RL based approach efficiently solves the highly complex joint optimization problem and improves total revenues for the virtual network operators.
In \cite{atallah2017reinforcement}, the downlink scheduling has been optimized for battery-charged roadside units in vehicular networks using RL methods to maximize the number of fulfilled service requests during a discharge period, where Q learning is employed to obtain the highest long-term returns.
The framework has been further extended in \cite{atallah2017deep}, where a deep RL based scheme has been proposed to learn a scheduling policy with high dimensional continuous inputs using end-to-end learning.
A distributed user association approach based on RL has been developed in \cite{li2017user} for vehicular networks with heterogeneous BSs. The proposed method leverages the $K$-armed bandit model to learn initial association for network load balancing and thereafter updates the solution directly using historical association patterns accumulated at each BS.
A similar handoff control problem in heterogeneous vehicular networks has been considered in \cite{xu2014fuzzy}, where a fuzzy Q-learning based approach has been proposed to always connect users to the best network without requirement of prior knowledge on handoff behavior.

\textcolor{black}{{This work differentiates itself from existing studies in at least two aspects. First, we explicitly model and solve the problem of improving the V2V payload delivery rate, i.e., the success probability of delivering packets of size $B$ within a time budget of $T$. This directly translates to reliability guarantee for periodic message sharing of V2V links, which is essentially a sequential decision making problem across many time steps within the message generation period.
Second, we propose a multi-agent RL based approach in this work to encourage and exploit V2V link cooperation to improve network level performance even when all V2V links make distributed spectrum access decisions based on local information. }}



\subsection{\textcolor{black}{{Contribution}}}
\textcolor{black}{{In this paper, we consider the spectrum sharing problem in high mobility vehicular networks, where multiple V2V links attempt to share the frequency spectrum preoccupied by V2I links.
To support diverse service requirements in vehicular networks, we design V2V spectrum and power allocation schemes that simultaneously maximize the capacity of V2I links for high bandwidth content delivery and meanwhile improve the payload delivery reliability of V2V links for periodic safety-critical message sharing.
The major contributions of this work are summarized as follows.}}
\begin{itemize}
\item \textcolor{black}{{We model the spectrum access of the multiple V2V links as a multi-agent problem and exploit recent progress of multi-agent RL \mbox{\cite{Omidshafiei2017deep,Foerster2017stabilising}} to develop a distributed spectrum and power allocation algorithm that simultaneously improves performance of both V2I and V2V links.}}
\item \textcolor{black}{{We provide a direct treatment of reliability guarantee for periodic safety message sharing of V2V links that adjusts V2V spectrum sub-band selection and power control in response to small-scale channel fading within the message generation period.}}
\item \textcolor{black}{{We show that through a proper reward design and training mechanism, the V2V transmitters can learn from interactions with the communication environment and figure out a clever strategy of working cooperatively with each other in a distributed way to optimize system level performance based on local information.}}
\end{itemize}


\subsection{Paper Organization}
The rest of the paper is organized as follows.
The system model is described in Section~\ref{marl:sec:sys}.
We present the proposed multi-agent RL based V2X resource sharing design in Section~\ref{marl:sec:resource}.
Section~\ref{marl:sec:simulation} provides our experimental results and concluding remarks are finally made in Section~\ref{marl:sec:conclusion}.

\section{System Model}\label{marl:sec:sys}

\textcolor{black}{{We consider a cellular based vehicular communication network in Fig.~\mbox{\ref{marl:fig:v2x}} with $M$ V2I and $K$ V2V links that provides simultaneous support for mobile high data rate entertainment and reliable periodic safety message sharing for advanced driving service, as discussed in 3GPP Release 15 for cellular V2X enhancement \mbox{\cite{3GPPr15v2x}}.
The V2I links leverage cellular (Uu) interfaces to connect $M$ vehicles to the BS for high data rate services while
the $K$ V2V links disseminate periodically generated safety messages via sidelink (PC5) interfaces with localized D2D communications.
We assume all transceivers use a single antenna and the set of V2I links and V2V links in the studied vehicular network are denoted by ${\mathcal M} = \{1,\cdots,M\}$ and ${\mathcal K} = \{1,\cdots,K\}$, respectively.
}}

\textcolor{black}{{We focus on Mode 4 defined in the cellular V2X architecture, where vehicles have a pool of radio resources that they can autonomously select for V2V communications \mbox{\cite{Molina2017LTEV}}. Such resource pools can overlap with that of the cellular V2I interfaces for better spectrum utilization provided necessary interference management design is in place, which is investigated in this work.
We further assume that the $M$ V2I links (uplink considered) have been preassigned orthogonal spectrum sub-bands with fixed transmission power, i.e., the $m$th V2I link occupies the $m$th sub-band.}}
As a result, the major challenge is to design an efficient spectrum sharing scheme for V2V links such that both V2I and V2V links achieve their respective goals with minimal signaling overhead given the strong dynamics underlying high mobility vehicular environments.

\textcolor{black}{{Orthogonal frequency division multiplexing (OFDM) is exploited to convert the frequency selective wireless channels into multiple parallel flat channels over different subcarriers. Several consecutive subcarriers are grouped to form a spectrum sub-band and we assume channel fading is approximately the same within one sub-band and independent across different sub-bands.
}}
During one coherence time period, the channel power gain, $g_k[m]$, of the $k$th V2V link over the $m$th sub-band (occupied by the $m$th V2I link) follows
\begin{align}
g_{k}[m] = \alpha_{k} h_{k}[m],
\end{align}
where $h_{k}[m]$ is the frequency dependent small-scale fading power component and assumed to be exponentially distributed with unit mean, and $\alpha_{k}$ captures the large-scale fading effect, including path loss and shadowing, assumed to be frequency independent.
The interfering channel from the $k'$th V2V transmitter to the $k$th V2V receiver over the $m$th sub-band, $g_{k',k}[m]$,
the interfering channel from the $k$th V2V transmitter to the BS over the $m$th sub-band, $g_{k,B}[m]$,
the channel from the $m$th V2I transmitter to the BS over the $m$th sub-band, $\hat{g}_{m,B}[m]$,
and the interfering channel from the $m$th V2I transmitter to the $k$th V2V receiver over the $m$th sub-band, $\hat{g}_{m,k}[m]$, are similarly defined.

The received signal-to-interference-plus-noise ratios (SINRs) of the $m$th V2I link and the $k$th V2V link over the $m$th sub-band are expressed as
\begin{align}\label{marl:eq:channel}
\gamma_{m}^c[m] = \frac{P_{m}^c \hat{g}_{m,B}[m]}{\sigma^2 + \sum\limits_{k}\rho_{k}[m] P_{k}^d[m] g_{k,B}[m]},
\end{align}
and
\begin{align}
\gamma_{k}^d[m] = \frac{P_{k}^d[m] g_{k}[m]}{\sigma^2 + I_k[m]},
\end{align}
respectively, where $P_{m}^c$ and $P_{k}^d[m]$ denote transmit powers of the $m$th V2I transmitter and the $k$th V2V transmitter over the $m$th sub-band, respectively, $\sigma^2$ is the noise power, and
\begin{align}\label{marl:eq:interference}
I_k[m] = P_{m}^c \hat{g}_{m,k}[m] + \sum\limits_{k'\ne k}\rho_{k'}[m] P_{k'}^d[m] g_{k',k}[m],
\end{align}
denotes the interference power. $\rho_{k}[m]$ is the binary spectrum allocation indicator with $\rho_{k}[m]=1$ implying the $k$th V2V link uses the $m$th sub-band and $\rho_{k}[m]=0$ otherwise.
We assume each V2V link only accesses one sub-band, i.e., $\sum\limits_m \rho_{k}[m] \le 1$.

Capacities of the $m$th V2I link and the $k$th V2V link over the $m$th sub-band are then obtained as
\begin{align}\label{eq:rateV2I}
C_m^c[m] = W\log(1+\gamma_m^c[m]),
\end{align}
and
\begin{align}
C_{k}^d[m] = W\log(1+\gamma_{k}^d[m]),
\end{align}
where $W$ is the bandwidth of each spectrum sub-band.

\textcolor{black}{{As described earlier, the V2I links are designed to support mobile high data rate entertainment services and hence an appropriate design objective is to maximize their sum capacity, defined as $\sum\limits_m C_m^c[m]$, for smooth mobile broadband access.
In the meantime, the V2V links are mainly responsible for reliable dissemination of safety-critical messages that are generated periodically with varying frequencies depending on vehicle mobility for advanced driving services. We mathematically model such a requirement as the delivery rate of packets of size $B$ within a time budget $T$ as
}}
\begin{align}\label{eq:V2Vdelivery}
\text{Pr}\left\{ \sum_{t=1}^T\sum\limits_{m=1}^M \rho_{k}[m] C_{k}^d[m,t] \ge B/\Delta_T\right\}, k\in\mathcal{K},
\end{align}
where $B$ denotes the size of the periodically generated V2V payload in bits, $\Delta_T$ is channel coherence time, and the index $t$ is added in $ C_{k}^d[m,t] $ to indicate the capacity of the $k$th V2V link at different coherence time slots.

\textcolor{black}{{To this end, the resource allocation problem investigated in this work is formally stated as: To design the V2V spectrum allocation, expressed through binary variables $\rho_k[m]$ for all $k\in\mathcal{K},m\in\mathcal{M}$, and the V2V transmission power, $P_k^d[m]$ for all $k\in\mathcal{K},m\in\mathcal{M}$, to simultaneously maximize the sum capacity of all V2I links $\sum\limits_m C_m^c[m]$ and the packet delivery rate of V2V links defined in \mbox{\eqref{eq:V2Vdelivery}}.
}}

\textcolor{black}{{High mobility in a vehicular environment precludes collection of accurate full CSI at a central controller, hence making distributed V2V resource allocation more preferable. Then how to coordinate actions of multiple V2V links such that they do not act selfishly in their own interests to compromise performance of the system as a whole remains challenging.
In addition, the packet delivery rate for V2V links, defined in \mbox{\eqref{eq:V2Vdelivery}}, involves sequential decision making across multiple coherence time slots within the time constraint $T$ and causes difficulties for conventional optimization methods due to exponential dimension increase.
To address these issues, we will exploit latest findings from multi-agent RL to develop a distributed algorithm for V2V spectrum access in the next section.
}}

\section{Multi-Agent RL Based Resource Allocation}\label{marl:sec:resource}

In the resource sharing scenario illustrated in Fig.~\ref{marl:fig:v2x}, multiple V2V links attempt to access limited spectrum occupied by V2I links, which can be modeled as a multi-agent RL problem.
Each V2V link acts as an agent and interacts with the unknown communication environment to gain experiences, which are then used to direct its own policy design.
Multiple V2V agents collectively explore the environment and refine spectrum allocation and power control strategies based on their own observations of the environment state.
While the resource sharing problem may appear a competitive game, we turn it into a fully cooperative one through using the same reward for all agents, in the interest of global network performance.

The proposed multi-agent RL based approach is divided into two phases, i.e., the learning (training) and the implementation phases.
We focus on settings with centralized learning and distributed implementation.
This means in the learning phase, the system performance-oriented reward is readily accessible to each individual V2V agent, which then adjusts its actions toward an optimal policy through updating its deep Q-network (DQN).
In the implementation phase, each V2V agent receives local observations of the environment and then selects an action according to its trained DQN on a time scale on par with the small-scale channel fading.
Key elements of the multi-agent RL based resource sharing design are described below in detail.



\subsection{State and Observation Space}

\begin{figure}[!t]
\centering
{\includegraphics[width=0.99\linewidth]{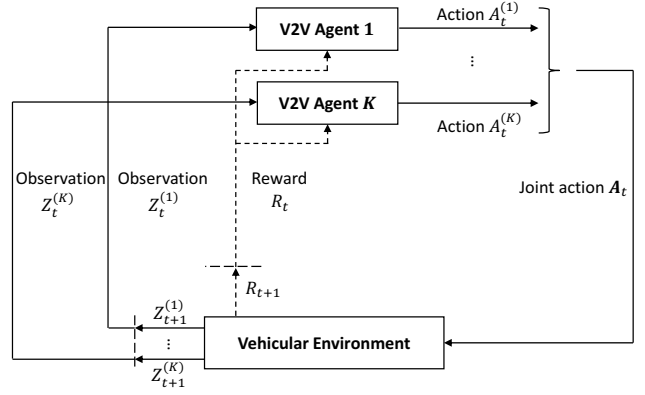}}
\caption{The agent-environment interaction in multi-agent RL formulation of the investigated resource sharing in vehicular networks.}\label{marl:fig:rl}
\end{figure}

\textcolor{black}{{In the multi-agent RL formulation of the resource sharing problem, each V2V link $k$ acts as an agent, concurrently exploring the unknown environment \mbox{\cite{Omidshafiei2017deep,Foerster2017stabilising}}.
Mathematically, the problem can be modeled as an MDP.
As shown in Fig.~\mbox{\ref{marl:fig:rl}}, at each coherence time step $t$, given the current environment state $S_{t}$, each V2V agent $k$ receives an observation $Z_{t}^{(k)}$ of the environment, determined by the observation function $O$ as $Z_{t}^{(k)} = O(S_t, k)$, and then takes an action $A_t^{(k)}$, forming a joint action $\mathbf{A}_t$. Thereafter, the agent receives a reward $R_{t+1}$ and the environment evolves to the next state $S_{t+1}$ with probability $p(s',r|s,\mathbf{a})$.
The new observations $Z_{t+1}^{(k)}$ are then received by each agent.
Please note that all V2V agents share the same reward in the system such that cooperative behavior among them is encouraged.}}

The true environment state, $S_t$, which could include global channel conditions and all agents' behaviors, is unknown to each individual V2V agent.
Each V2V agent can only acquire knowledge of the underlying environment through the lens of an observation function.
The observation space of an individual V2V agent $k$ contains local channel information, including
its own signal channel, $g_{k}[m]$, for all $m\in\mathcal{M}$,
interference channels from other V2V transmitters, $g_{k',k}[m]$, for all $k'\ne k$ and $m\in\mathcal{M}$,
the interference channel from its own transmitter to the BS, $g_{k,B}[m]$, for all $m\in\mathcal{M}$,
and
the interference channel from all V2I transmitters, $\hat{g}_{m,k}[m]$, for all $m\in\mathcal{M}$.
Such channel information, except $g_{k,B}[m]$, can be accurately estimated by the receiver of the $k$th V2V link at the beginning of each time slot $t$ and we assume it is also available instantaneously at the transmitter through delay-free feedback \cite{Nasir2018deep}.
The channel $g_{k,B}[m]$ is estimated at the BS in each time slot $t$ and then broadcast to all vehicles in its coverage, which incurs small signaling overhead.
The received interference power over all bands, $I_{k}[m]$, for all $m\in\mathcal{M}$, expressed in \eqref{marl:eq:interference}, can be measured at the V2V receiver and is also introduced in the local observation.
In addition, the local observation space includes
the remaining V2V payload, $B_{k}$,
and the remaining time budget, $T_{k}$, to better capture the queuing states of each V2V link.
As a result, the observation function for an agent $k$ is summarized as
\begin{align}
O(S_t, k) = \left\{ B_{k}, T_{k}, \{I_{k}[m]\}_{m\in\mathcal{M}}, \{G_k[m]\}_{m\in\mathcal{M}}\right\},
\end{align}
with $G_k[m] = \{g_{k}[m], g_{k',k}[m], g_{k,B}[m], \hat{g}_{m,k}[m]\}$.

Independent Q-learning \cite{tan1993multi} is among the most popular methods to solve multi-agent RL problems, where each agent learns a decentralized policy based on its own action and observation, treating other agents as part of the environment.
However, naively combining DQN with independent Q-learning is problematic since each agent would face a nonstationary environment while other agents are also learning to adjust their behaviors.
The issue grows even more severe with experience replay, which is the key to the success of DQN, in that sampled experiences no longer reflect current dynamics and thus destabilize learning.
To address this issue, we adopt the fingerprint-based method developed in \cite{Foerster2017stabilising}.
The idea is that while the action-value function of an agent is nonstationary with other agents changing their behaviors over time, it can be made stationary conditioned on other agents' policies.
This means we can augment each agent's observation space with an estimate of other agents' policies to avoid nonstationarity, which is the essential idea of hyper Q-learning \cite{tesauro2004extending}.
However, it is undesirable for the action-value function to include as input all parameters of other agents' neural networks, ${\boldsymbol{\theta}}_{-i}$, since the policy of each agent consists of a high dimensional DQN.
Instead, it is proposed in \cite{Foerster2017stabilising} to simply include a low-dimensional fingerprint that tracks the trajectory of the policy change of other agents.
This method works since nonstationarity of the action-value function results from changes of other agents' policies over time, as opposed to the policies themselves.
Further analysis reveals that each agent's policy change is highly correlated with the training iteration number $e$ as well as its rate of exploration, e.g., the probability of random action selection, $\epsilon$, in the $\epsilon$-greedy policy widely used in Q-learning.
Therefore, we include both of them in the observation for an agent $k$, expressed as
\begin{align}\label{eq:observation}
Z_t^{(k)} = \left\{O(S_t, k), e, \epsilon\right\}.
\end{align}

\subsection{Action Space}
The resource sharing design of vehicular links comes down to the spectrum sub-band selection and transmission power control for V2V links. While the spectrum naturally breaks into $M$ disjoint sub-bands, each preoccupied by one V2I link, the V2V transmission power typically takes continuous value in most existing power control literature. In this paper, however, we limit the power control options to four levels, i.e., $[23, 10, 5, -100]$ dBm, for the sake of both ease of learning and practical circuit restriction.
It is noted that the choice of $-100$ dBm effectively means zero V2V transmission power.
As a result, the dimension of the action space is $4\times M$, with each action corresponding to one particular combination of a spectrum sub-band and power selection.

\subsection{Reward Design}\label{sec:reward}
What makes RL particularly appealing for solving problems with hard-to-optimize objectives is the flexibility in its reward design.
The system performance can be improved when the designed reward signal at each step correlates with the desired objective.
In the investigated V2X spectrum sharing problem described in Section~\ref{marl:sec:sys}, our objectives are twofold: Maximizing the sum V2I capacity while increasing the success probability of V2V payload delivery within a certain time constraint $T$.

In response to the first objective, we simply include the instantaneous sum capacity of all V2I links, $\sum\limits_{m\in\mathcal{M}} C_m^c[m, t]$ as defined in \eqref{eq:rateV2I}, in the reward at each time step $t$.
To achieve the second objective, for each agent $k$, we set the reward $L_k$ equal to the effective V2V transmission rate until the payload is delivered, after which the reward is set to a constant number, $\beta$, that is greater than the largest possible V2V transmission rate.
As such, the V2V-related reward at each time step $t$ is set as
\begin{align}\label{eq:v2vReward}
L_k(t) =
\begin{cases}
\sum\limits_{m=1}^{M}\rho_k[m] C_k^d[m,t], & \text{if $B_k \ge 0$},\\
\beta, & \text{otherwise}.
\end{cases}
\end{align}

\textcolor{black}{{The goal of learning is to find an optimal policy $\pi_*$ (a mapping from states in $\mathcal{S}$ to probabilities of selecting each action in $\mathcal{A}$) that maximizes the expected return from any initial state $s$, where the return, denoted by $G_t$, is defined as the cumulative discounted rewards with a discount rate $\gamma$, i.e., }}
\begin{align}\label{eq:G}
G_t = \sum\limits_{k=0}^{\infty} \gamma^k R_{t+k+1}, 0 \le \gamma \le 1.
\end{align}
We observe that if setting the discount rate $\gamma$ to 1, greater cumulative rewards translate to a larger amount of transmitted data for V2V links until the payload delivery is finished.
Hence maximizing the expected cumulative rewards encourages more data to be delivered for V2V links when the remaining payload is still nonzero, i.e., $B_k \ge 0$.
In addition, the learning process also attempts to achieve as many rewards of $\beta$ as possible, effectively leading to higher possibility of successful delivery of V2V payload.

In practice, $\beta$ is a hyperparameter that needs to be tuned empirically.
\textcolor{black}{{In our training, $\beta$ is tuned such that it is greater than the largest V2V transmission rate that is obtained from running a few steps of random resource allocation, but should not be ``too big'' and ideally less than twice of the largest value from our tuning experience.
The design of $\beta$ represents our thinking about the tradeoff between designing the reward purely toward the ultimate goal and learning efficiency.
For pure goal-directed consideration, we just set the rewards at each step to $0$ until the V2V payload is delivered, beyond which point the reward is set to $1$. However, our tuning experience suggests such a design will hinder the learning process since the agent can hardly learn anything useful at the beginning of each episode as it always receives a reward of $0$ for this period. We then impart some prior knowledge into the reward, i.e., higher V2V transmission rates should be helpful in improving V2V payload delivery rate. Hence, we come up with the reward design described in \mbox{\eqref{eq:v2vReward}} to blend this two extremes of reward designs.
}}

To this end, we set the reward at each time step $t$ as
\begin{align}
R_{t+1} = \lambda_c \sum\limits_{m} C_m^c[m,t] + \lambda_d \sum\limits_{k} L_k(t),
\end{align}
where $\lambda_c$ and $\lambda_d$ are positive weights to balance V2I and V2V objectives.

\subsection{Learning Algorithm}\label{sec:algorithm}

\begin{table}[!t]
\begin{algorithm}[H]
\caption{Resource Sharing with Multi-Agent RL} \label{marl:algm:randomized}
\begin{algorithmic}[1]
\normalsize{
\STATE Start environment simulator, generating vehicles and links
\STATE Initialize Q-networks for all agents randomly
\FOR{each episode}
\STATE Update vehicle locations and large-scale fading $\boldsymbol{\alpha}$
\STATE Reset $B_k=B$ and $T_k=T$, for all $k\in{\mathcal{K}}$
\FOR{each step $t$}
\FOR{each V2V agent $k$}
\STATE Observe $Z_t^{(k)}$
\STATE Choose action $A_t^{(k)}$ from $Z_t^{(k)}$ according to $\epsilon$-greedy policy
\ENDFOR
\STATE All agents take actions and receive reward $R_{t+1}$
\STATE Update channel small-scale fading
\FOR{each V2V agent $k$}
\STATE Observe  $Z_{t+1}^{(k)}$
\STATE Store $\left(Z_t^{(k)}, A_t^{(k)}, R_{t+1}, Z_{t+1}^{(k)}\right)$ in the replay memory $\mathcal{D}_k$
\ENDFOR
\ENDFOR
\FOR{each V2V agent $k$}
\STATE Uniformly sample mini-batches from $\mathcal{D}_k$
\STATE Optimize error between Q-network and learning targets, defined in \eqref{marl:eq:dqn}, using variant of stochastic gradient descent
\ENDFOR
\ENDFOR
}
\end{algorithmic}
\end{algorithm}
\end{table}

We focus on an episodic setting with each episode spanning the V2V payload delivery time constraint $T$.
Each episode starts with a randomly initialized environment state (determined by the initial transmission powers of all vehicular links, channel states, etc.) and a full V2V payload of size $B$ for transmission, and lasts until the end of $T$.
The change of small-scale channel fading triggers a transition of the environment state and causes each individual V2V agent to adjust its action.

\subsubsection{\textcolor{black}{{Training Procedure}}}
\textcolor{black}{{We leverage deep Q-learning with experience replay \mbox{\cite{mnih2015human}} to train the multiple V2V agents for effective learning of spectrum access policies. Q-Learning \mbox{\cite{watkins1992q}} is based on the concept of action-value function, $q_{\pi}(s,a)$, for policy $\pi$, which is defined as the expected return starting from the state $s$, taking the action $a$, and thereafter following the policy $\pi$, formally expressed as}}
\begin{align}
q_{\pi}(s,a) = \mathbb{E}_{\pi}\left[G_t | S_t = s, A_t = a \right],
\end{align}
\textcolor{black}{{where $G_t$ is defined in \mbox{\eqref{eq:G}}.
It is easy to determine the optimal policy once its action-value function, $q_*(s,a)$, is obtained.
It has been shown in \mbox{\cite{sutton1998reinforcement}} that with a variant of the stochastic approximation condition on the learning rate and the assumption that all state-action pairs continue to be updated, the learned action-value function in Q-learning converges with probability $1$ to the optimal $q_*$. In deep Q-learning \mbox{\cite{mnih2015human}}, a deep neural network parameterized by $\theta$, called DQN, is used to represent the action-value function.}}

\textcolor{black}{{Each V2V agent $k$ has a dedicated DQN that takes as input the current observation $Z_t^{(k)}$ and outputs the value functions corresponding to all actions.
We train the Q-networks through running multiple episodes and, at each training step, all V2V agents explore the state-action space with some soft policies, e.g., $\epsilon$-greedy, meaning that the action with maximal estimated value is chosen with probability $1-\epsilon$ while a random action is instead selected with probability $\epsilon$.
Following the environment transition due to channel evolution and actions taken by all V2V agents, each agent $k$ collects and stores the transition tuple, $\left(Z_t^{(k)}, A_t^{(k)}, R_{t+1}, Z_{t+1}^{(k)}\right)$, in a replay memory.
}}
\textcolor{black}{{
At each episode, a mini-batch of experiences $\mathcal{D}$ are uniformly sampled from the memory for updating $\theta$ with variants of stochastic gradient-descent methods, hence the name experience replay, to minimize the sum-squared error:}}
\begin{align}\label{marl:eq:dqn}
\sum\limits_{\mathcal{D}} \left[R_{t+1} + \gamma\max\limits_{a'}Q(Z_{t+1}, a';\theta^-) - Q(Z_t, A_t; \theta) \right]^2,
\end{align}
\textcolor{black}{{where $\theta^-$ is the parameter set of a target Q-network, which are duplicated from the training Q-network parameter set $\theta$ periodically and fixed for a couple of updates.
Experience replay improves sample efficiency through repeatedly sampling stored experiences and breaks correlation in successive updates, thus also stabilizing learning.
The training procedure is summarized in Algorithm~\mbox{\ref{marl:algm:randomized}}.
}}

\subsubsection{Distributed Implementation}
During the implementation phase, at each time step $t$, each V2V agent $k$ estimates local channels and compiles a local observation, $Z^{(k)}_t$, of the environment based on \eqref{eq:observation} with $e$ and $\epsilon$ set to the values from the very last training step.
Then it selects an action, $A_t^{(k)}$, with the maximum action value according to its trained Q-network.
Afterwards, all V2V links start transmission with the power level and frequency spectrum sub-band determined by their selected actions.

Note that the computation intensive training procedure in Algorithm~\ref{marl:algm:randomized} can be performed offline for many episodes over different channel conditions and network topology changes while the inexpensive implementation procedure is executed online for network deployment.
The trained DQNs for all agents only need to be updated when the environment characteristics have experienced significant changes, say, once a week or even a month, depending on environment dynamics and network performance requirements.

\section{Simulation Results}\label{marl:sec:simulation}

\begin{table}[!t]
\centering
\caption{Simulation Parameters \cite{3GPPr14v2x,3GPPsimulation}}\label{marl:table:simulaton}
\begin{threeparttable}
\begin{tabular}{|m{0.45\linewidth}|m{0.4\linewidth}|}
\hline
{\textbf{Parameter}} & \textbf{Value} \\\hline
Number of V2I links $M$ & 4\\ \hline
Number of V2V links $K$ & 4\\ \hline
Carrier frequency & 2 GHz \\\hline
Bandwidth & 4 MHz \\\hline
BS antenna height & 25 m \\\hline
BS antenna gain & 8 dBi \\ \hline
BS receiver noise figure & 5 dB \\ \hline
Vehicle antenna height & 1.5 m \\\hline
Vehicle antenna gain & 3 dBi \\\hline
Vehicle receiver noise figure & 9 dB \\\hline
Absolute vehicle speed $v$ & 36 km/h \\\hline
Vehicle drop and mobility model & Urban case of A.1.2 in \cite{3GPPr14v2x}\tnote{*}\\ \hline
V2I transmit power $P^c$ & 23 dBm \\ \hline
V2V transmit power $P^d$ & [23,10,5,-100] dBm \\ \hline
Noise power $\sigma^2$  & -114 dBm \\ \hline
Time constraint of V2V payload transmission $T$ & 100 ms \\ \hline
V2V payload size $B$ & $[1,2,\cdots]\times$ 1060 bytes \\\hline
\end{tabular}
\begin{tablenotes}
\item[*] We shrink the height and width of the simulation area by a factor of 2.
\end{tablenotes}
\end{threeparttable}
\end{table}

\begin{table}[!t]
\centering
\caption{Channel Models for V2I and V2V Links \cite{3GPPr14v2x}}\label{marl:table:channel}
\begin{tabular}{|m{0.3\linewidth}|m{0.28\linewidth}|m{0.24\linewidth}|}
\hline
\textbf{Parameter} & \textbf{V2I Link} & \textbf{V2V Link} \\ \hline
Path loss model & 128.1 + 37.6$\log_{10}d$, $d$ in km  & LOS in WINNER + B1 Manhattan \cite{WINNER}\\ \hline
Shadowing distribution & Log-normal & Log-normal\\ \hline
Shadowing standard deviation $\xi$ & 8 dB & 3 dB \\ \hline
Decorrelation distance & 50 m & 10 m \\ \hline
Path loss and shadowing update & A.1.4 in \cite{3GPPr14v2x} every 100 ms & A.1.4 in \cite{3GPPr14v2x} every 100 ms \\\hline
Fast fading & Rayleigh fading & Rayleigh fading \\ \hline
Fast fading update & Every 1 ms & Every 1 ms \\\hline
\end{tabular}
\end{table}

\textcolor{black}{{In this section, simulation results are presented to validate the proposed multi-agent RL based resource sharing scheme for vehicular networks.
We custom built our simulator following the evaluation methodology for the urban case defined in Annex A of 3GPP TR~36.885 \mbox{\cite{3GPPr14v2x}}, which describes in detail vehicle drop models, densities, speeds, direction of movement, vehicular channels, V2V data traffic, etc..
The $M$ V2I links are started by $M$ vehicles and the $K$ V2V links are formed between each vehicle with its surrounding neighbors.
Major simulation parameters are listed in Table~\mbox{\ref{marl:table:simulaton}} and the channel models for V2I and V2V links are described in Table~\mbox{\ref{marl:table:channel}}.
Note that all parameters are set to the values specified in Tables~\mbox{\ref{marl:table:simulaton}} and \mbox{\ref{marl:table:channel}} by default, whereas the settings in each figure take precedence wherever applicable.
}}

The DQN for each V2V agent consists of $3$ fully connected hidden layers, containing $500, 250$, and $120$ neurons, respectively. The rectified linear unit (ReLU), $f(x) = \max(0,x)$, is used as the activation function and RMSProp optimizer \cite{ruder2016overview} is used to update network parameters with a learning rate of $0.001$.
We train each agent's Q-network for a total of $3,000$ episodes and the exploration rate $\epsilon$ is linearly annealed from $1$ to $0.02$ over the beginning $2,400$ episodes and remains constant afterwards.
It is noted that we fix the large-scale fading for a couple of training episodes and let the small-scale fading change over each step such that the learning algorithm can better acquire the underlying fading dynamics, thus helping stabilise training.
In addition, we fix the V2V payload size $B$ in the training stage to be of $2\times 1060$ bytes, but vary the sizes in the testing stage to verify robustness of the proposed method.

\begin{figure}
\centering
\includegraphics[width=0.95\linewidth]{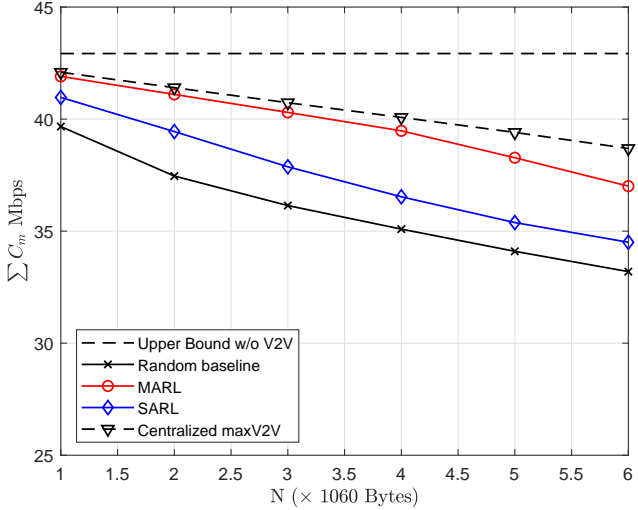}
\caption{Sum capacity performance of V2I links with varying V2V payload sizes $B$.}\label{marl:fig:v2i}
\end{figure}

\begin{figure}[t]
\centering
\includegraphics[width=0.95\linewidth]{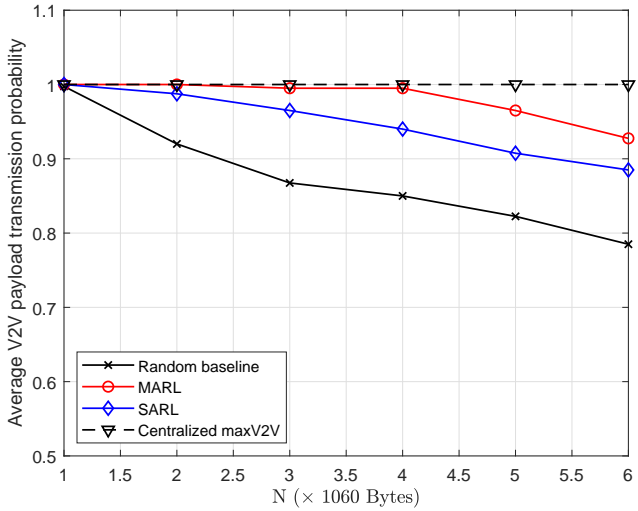}
\caption{V2V payload transmission success probability with varying payload sizes $B$.}\label{marl:fig:v2v}
\end{figure}

\textcolor{black}{{We compare in Figs.~\mbox{\ref{marl:fig:v2i} and \ref{marl:fig:v2v}} the proposed multi-agent RL based resource sharing scheme, termed MARL, against the following two baseline methods that are executed in a distributed manner.}}
\begin{enumerate}
    \item \textcolor{black}{{The single-agent RL based algorithm in \mbox{\cite{Ye2019deep}}, termed SARL, where at each moment only one V2V agent updates its action, i.e., spectrum sub-band selection and power control, based on locally acquired information and a trained DQN  while others agents' actions remain unchanged. A single DQN is shared across all V2V agents. }}
    \item \textcolor{black}{{The random baseline, which chooses the spectrum sub-band and transmission power for each V2V link in a random fashion at each time step. }}
\end{enumerate}

\textcolor{black}{{We further benchmark the proposed MARL method in Algorithm~\mbox{\ref{marl:algm:randomized}} against the theoretical performance upper bounds of the V2I and V2V links, derived from the following two idealistic (and extreme) schemes.}}
\begin{enumerate}
    \item \textcolor{black}{{We disable the transmission of all V2V links to obtain the upper bound of V2I performance, hence the name upper bound without V2V. In this case, the packet delivery rates for all V2V links are exactly zero, thus not shown in Fig.~\mbox{\ref{marl:fig:v2v}}.}}
    \item \textcolor{black}{{We exclusively focus on improving V2V performance while ignoring the requirement of V2I links. Such an assumption breaks the sequential decision making of delivering $B$ bytes over multiple steps within the time constraint $T$ into separate optimization of sum V2V rates over each step. Then, we exhaustively search the action space of all $K$ V2V agents in each step to maximize sum V2V rates. Apart from the complexity due to exhaustive search, this scheme needs to be performed in a centralized way with accurate global CSI available, hence the name centralized maxV2V.}}
\end{enumerate}
\textcolor{black}{{We remark that although these two schemes are way too idealistic and cannot be implemented in practice, they provide meaningful performance upper bounds for V2I and V2V links that illustrate how closely the proposed method can approach the limit. }}

Fig.~\ref{marl:fig:v2i} shows the V2I performance with respect to increasing V2V payload sizes $B$ for different resource sharing designs.
From the figure, the performance drops for all schemes (except the upper bound) with growing V2V payload sizes.
An increase of V2V payload leads to longer V2V transmission duration and possibly higher V2V transmit power in order to improve V2V payload transmission success probability.
This will inevitably cause stronger interference to V2I links for a longer period and thus jeopardize their capacity performance.
We observe that the proposed MARL method in Algorithm~\ref{marl:algm:randomized} achieves better performance than the other two baseline schemes across different V2V payload sizes although it is trained with a fixed size of $2\times 1060$ bytes, demonstrating its robustness against V2V payload variation.
\textcolor{black}{{It performs measurably close to the V2I performance upper bound, within $14\%$ degradation even in the worst case of $6 \times 1060$ bytes of payload.
We also note that the centralized maxV2V scheme attains remarkable performance in terms of V2I performance.  This could be due to the packet delivery rates of V2V links have been substantially enhanced with centralized maxV2V and the V2V links incur no interference to V2I links once their payload delivery has finished. This is an interesting observation that warrants further investigation into the performance tradeoff between V2I and V2V links.
That said, the proposed distributed MARL method tightly follows the idealistic centralized maxV2V scheme, further demonstrating its effectiveness.
}}

Fig.~\ref{marl:fig:v2v} shows the success probability of V2V payload delivery against growing payload sizes $B$ under different spectrum sharing schemes.
\textcolor{black}{{From the figure, as the V2V payload size grows larger, the transmission success probabilities drop for all three distributed algorithms, including the proposed MARL, while the centralized maxV2V can achieve $100\%$ packet delivery throughout the tested cases.
The proposed MARL method achieves significantly better performance than the two baseline distributed methods and stays very close to the centralized maxV2V scheme.
Remarkably, the proposed method attains $100\%$ V2V payload delivery probability for $B=1060$ and $B=2\times 1060$ bytes and achieves close to perfect performance for $B=3\times 1060$ and $B=4\times 1060$ bytes.}}

We also observe from Fig.~\ref{marl:fig:v2v} that the proposed MARL method achieves highly desirable V2V performance for the low payload cases and suffers from noticeable degradation when the payload size grows beyond $4\times 1060$ bytes.
In conjunction with the observations from Fig.~\ref{marl:fig:v2i}, we conclude that the robustness of the proposed multi-agent RL based method against V2V payload variation should be taken with a grain of salt: Within a reasonable region of payload size change, the trained DQN is good, which, however, needs to be updated if the change grows beyond the acceptable margin.
\textcolor{black}{{However, the exact range of such acceptable margin is difficult to be determined in general, which would depend on the actual system parameter settings.
For the current setting, we can conclude that no noticeable performance loss is spotted when the packet size is no greater than $4\times1060$ bytes and to maintain a V2V delivery rate above $95\%$, the packet size needs to be  no larger than $5\times 1060$ bytes.
Again, such observations are based on the particular setting for the simulation and extra caution is needed when generalizing them.
That said, we can still validate the advantage of the proposed spectrum access design since it outperforms the other two distributed baselines even in the untrained scenarios.
}}


\begin{figure}[t]
\centering
\includegraphics[width=0.95\linewidth]{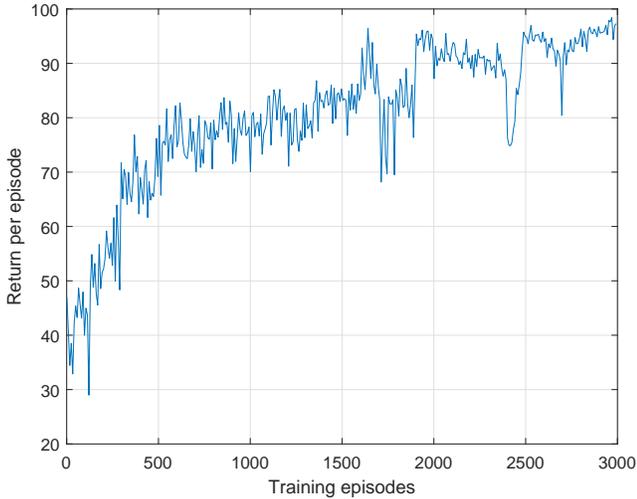}
\caption{Return for each training episode with increasing iterations. The V2V payload size $B = 2,120$ bytes.}\label{marl:fig:reward}
\end{figure}

We show in Fig.~\ref{marl:fig:reward} the cumulative rewards per training episode with increasing training iterations to study the convergence behavior of the proposed multi-agent RL method.
From the figure, the cumulative rewards per episode improve as training continues, demonstrating the effectiveness of the proposed training algorithm.
When the training episode approximately reaches $2,000$, the performance gradually converges despite some fluctuations due to mobility-induced channel fading in vehicular environments.
Based on such an observation, we train each agent's Q-network for $3,000$ episodes when evaluating the performance of V2I and V2V links in Figs.~\ref{marl:fig:v2i} and \ref{marl:fig:v2v}, which should provide a safe convergence guarantee.

\begin{figure}
\centering
\subfigure[The remaining payload of MARL.]{\includegraphics[clip,width=0.95\linewidth]{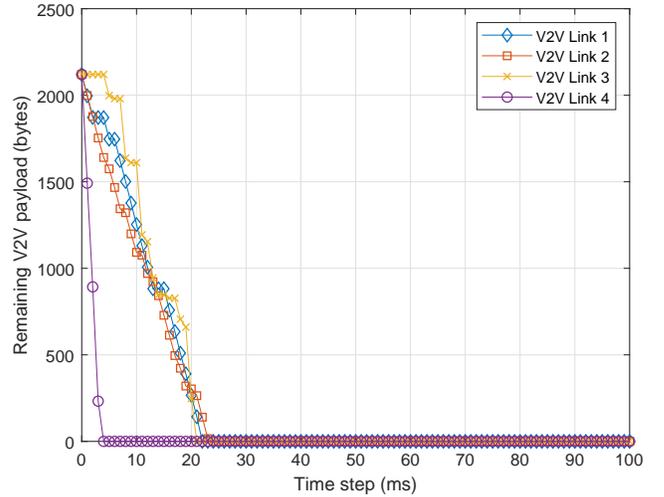}}
\subfigure[The remaining payload of the random baseline.]{\includegraphics[clip,width=0.95\linewidth]{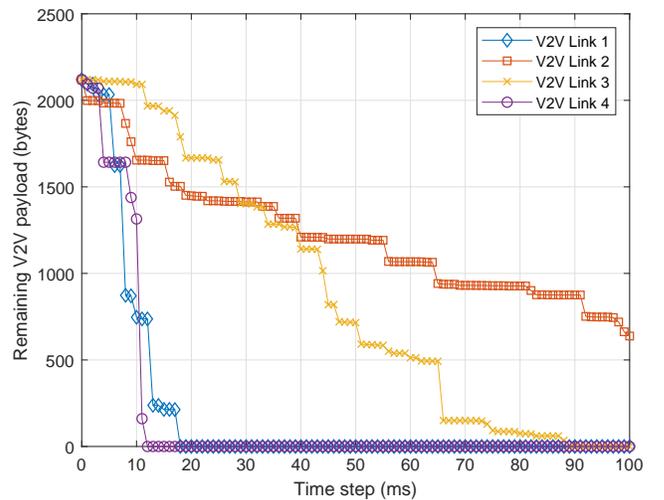}}
\caption{The change of the remaining V2V payload of the proposed MARL and the random baseline resource sharing schemes within the time constraint $T=100$ ms. The initial payload size $B = 2,120$ byte.}\label{fig:payload}
\end{figure}

To understand why the proposed multi-agent RL based method achieves better performance compared with the random baseline, we select an episode in which the proposed method enables all V2V links to successfully deliver the payload of $2,120$ bytes while the random baseline fails.
We plot in Fig.~\ref{fig:payload} the change of the remaining V2V payload within the time constraint, i.e., $T=100$ ms, for all V2V links.
From Fig.~\ref{fig:payload}(a), the V2V Link 4 finishes payload delivery early in the episode while the other three links end transmission roughly at the same time for the proposed multi-agent RL based method.
For the random baseline, Fig.~\ref{fig:payload}(b) shows that V2V Links 1 and 4 successfully deliver all payload early in the episode. V2V Link 3 also finishes payload transmission albeit much later in the episode while V2V Link 2 fails to deliver the required payload.

\begin{figure}[t]
\centering
\subfigure[V2V transmission rates of MARL.]{\includegraphics[clip,width=0.95\linewidth]{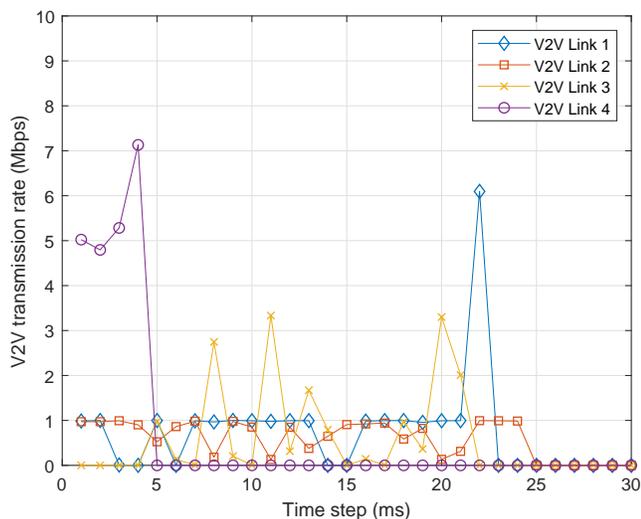}}
\subfigure[V2V transmission rates of the random baseline.]{\includegraphics[clip,width=0.95\linewidth]{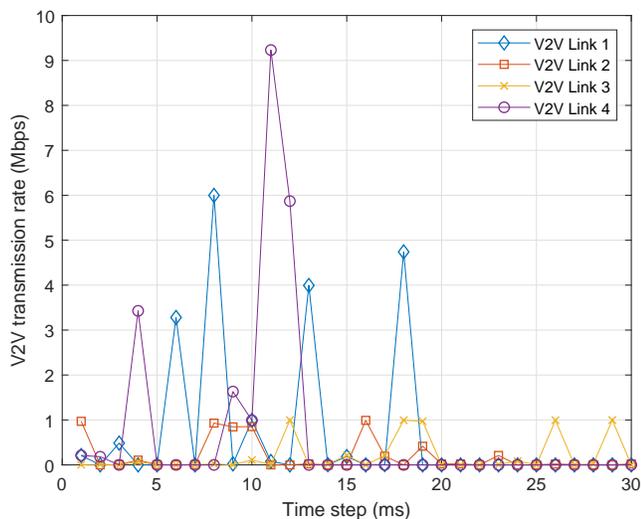}}
\caption{V2V transmission rates of the proposed MARL and the random baseline resource allocation schemes within the same episode as Fig.~\ref{fig:payload}. Only the results of the beginning $30$ ms are plotted for better presentation. The initial payload size $B = 2,120$ bytes.}\label{fig:v2vRate}
\end{figure}

In Fig.~\ref{fig:v2vRate}, we further show the instantaneous rates of all V2V links under the two different resource allocation schemes at each step in the same episode as Fig.~\ref{fig:payload}.
Several valuable observations can be made from comparing Figs.~\ref{fig:v2vRate}(a) and (b) that demonstrate the effectiveness of the proposed method in encouraging cooperation among multiple V2V agents.
From Fig.~\ref{fig:v2vRate}(a), with the proposed method, V2V Link 4 gets very high transmission rates at the beginning to finish transmission early such that the good channel condition of this link is fully exploited and no interference will be generated toward other links at later stages of the episode.
V2V Link 1 keeps low transmission rates at first such that the vulnerable V2V Links 2 and 3 can get relatively good transmission rates to deliver payload, and then jumps to high data rates to deliver its own data when Links 2 and 3 almost finish transmission.
Moreover, a closer examination of the rates of Links 2 and 3 reveals that the two links figure out a clever strategy to take turns to transmit such that both of their payloads can be delivered quickly.
To summarize, the proposed multi-agent RL based method learns to leverage good channels of some V2V links and meanwhile provides protection for those with bad channel conditions.
The success probability of V2V payload transmission is thus significantly improved.
In contrast, Fig.~\ref{fig:v2vRate}(b) shows that the random baseline method fails to provide such protection for vulnerable V2V links, leading to high probability of failed payload delivery for them.

\section{Conclusion}\label{marl:sec:conclusion}
In this paper, we have developed a distributed resource sharing scheme based on multi-agent RL for vehicular networks with multiple V2V links reusing the spectrum of V2I links.
A fingerprint-based method has been exploited to address nonstationary issues of independent Q-learning for multi-agent RL problems when combined with DQN with experience replay.
The proposed multi-agent RL based method is divided into a centralized training stage and a distributed implementation stage.
We demonstrate that through such a mechanism, the proposed resource sharing scheme is effective in encouraging cooperation among V2V links to improve system level performance although decision making is performed locally at each V2V transmitter.
Future work will include an in-depth analysis and comparison of the robustness of both single-agent and multi-agent RL based algorithms to gain better understanding on when the trained Q-networks need to be updated and how to efficiently perform such updates.
\textcolor{black}{{Extension of the proposed multi-agent RL based resource allocation method to the multiple-input multiple-output (MIMO) and the millimeter MIMO scenarios for vehicular communications is also an interesting direction worth further investigation.}}


\bibliographystyle{IEEEtran}
\bibliography{marl_v2x}

\begin{thebibliography}{10}
\providecommand{\url}[1]{#1}
\csname url@samestyle\endcsname
\providecommand{\newblock}{\relax}
\providecommand{\bibinfo}[2]{#2}
\providecommand{\BIBentrySTDinterwordspacing}{\spaceskip=0pt\relax}
\providecommand{\BIBentryALTinterwordstretchfactor}{4}
\providecommand{\BIBentryALTinterwordspacing}{\spaceskip=\fontdimen2\font plus
\BIBentryALTinterwordstretchfactor\fontdimen3\font minus
  \fontdimen4\font\relax}
\providecommand{\BIBforeignlanguage}[2]{{%
\expandafter\ifx\csname l@#1\endcsname\relax
\typeout{** WARNING: IEEEtran.bst: No hyphenation pattern has been}%
\typeout{** loaded for the language `#1'. Using the pattern for}%
\typeout{** the default language instead.}%
\else
\language=\csname l@#1\endcsname
\fi
#2}}
\providecommand{\BIBdecl}{\relax}
\BIBdecl

\bibitem{Liang2017vehicular}
L.~Liang, H.~Peng, G.~Y. Li, and X.~Shen, ``{Vehicular communications: A
  physical layer perspective},'' \emph{IEEE Trans. Veh. Technol.}, vol.~66,
  no.~12, pp. 10\,647--10\,659, Dec. 2017.

\bibitem{Peng2018vehicular}
H.~Peng, L.~Liang, X.~Shen, and G.~Y. Li, ``Vehicular communications: A network
  layer perspective,'' \emph{IEEE Trans. Veh. Technol.}, vol.~2, no.~68, pp.
  1064--1078, Feb. 2019.

\bibitem{3GPPr14v2x}
\emph{{3rd Generation Partnership Project; Technical Specification Group Radio
  Access Network; Study on LTE-based V2X Services; (Release 14)}}, 3GPP TR
  36.885 V14.0.0, Jun. 2016.

\bibitem{3GPPr15v2x}
\emph{{3rd Generation Partnership Project; Technical Specification Group Radio
  Access Network; Study on enhancement of 3GPP Support for 5G V2X Services;
  (Release 15)}}, 3GPP TR 22.886 V15.1.0, Mar. 2017.

\bibitem{Molina2017LTEV}
R.~Molina-Masegosa and J.~Gozalvez, ``{LTE-V for sidelink 5G V2X vehicular
  communications: A new 5G technology for short-range vehicle-to-everything
  communications},'' \emph{IEEE Veh. Technol. Mag.}, vol.~12, no.~4, pp.
  30--39, Dec. 2017.

\bibitem{sutton1998reinforcement}
R.~S. Sutton, A.~G. Barto \emph{et~al.}, \emph{Reinforcement learning: An
  introduction}.\hskip 1em plus 0.5em minus 0.4em\relax MIT press, 1998.

\bibitem{mnih2015human}
V.~Mnih, K.~Kavukcuoglu, D.~Silver, A.~A. Rusu, J.~Veness, M.~G. Bellemare,
  A.~Graves, M.~Riedmiller, A.~K. Fidjeland, G.~Ostrovski \emph{et~al.},
  ``Human-level control through deep reinforcement learning,'' \emph{Nature},
  vol. 518, no. 7540, pp. 529--533, Feb. 2015.

\bibitem{silver2016mastering}
D.~Silver, A.~Huang, C.~J. Maddison, A.~Guez, L.~Sifre, G.~Van Den~Driessche,
  J.~Schrittwieser, I.~Antonoglou, V.~Panneershelvam, M.~Lanctot \emph{et~al.},
  ``Mastering the game of go with deep neural networks and tree search,''
  \emph{Nature}, vol. 529, no. 7587, pp. 484--489, Jan. 2016.

\bibitem{He2017deep}
Y.~He, Z.~Zhang, F.~R. Yu, N.~Zhao, H.~Yin, V.~C.~M. Leung, and Y.~Zhang,
  ``Deep-reinforcement-learning-based optimization for cache-enabled
  opportunistic interference alignment wireless networks,'' \emph{IEEE Trans.
  Veh. Technol.}, vol.~66, no.~11, pp. 10\,433--10\,445, Nov. 2017.

\bibitem{He2018secure}
Y.~He, F.~R. Yu, N.~Zhao, and H.~Yin, ``{Secure social networks in 5G systems
  with mobile edge computing, caching, and device-to-device communications},''
  \emph{IEEE Wireless Commun.}, vol.~25, no.~3, pp. 103--109, Jun. 2018.

\bibitem{Mao2016resource}
H.~Mao, M.~Alizadeh, I.~Menache, and S.~Kandula, ``{Resource management with
  deep reinforcement learning},'' in \emph{Proc. ACM Workshop Hot Topics Netw.
  (HotNets)}, 2016, pp. 50--56.

\bibitem{Ye2019deep}
H.~Ye, G.~Y. Li, and B.-H. Juang, ``{Deep reinforcement learning based resource
  allocation for V2V communications},'' \emph{IEEE Trans. Veh. Technol.},
  vol.~68, no.~4, pp. 3163–--3173, Apr. 2019.

\bibitem{Botsov2014location}
M.~Botsov, M.~Kl{\"{u}}gel, W.~Kellerer, and P.~Fertl, ``{Location dependent
  resource allocation for mobile device-to-device communications},'' in
  \emph{Proc. IEEE WCNC}, Apr. 2014, pp. 1679--1684.

\bibitem{Sun2016radio}
W.~Sun, E.~G. Str{\"o}m, F.~Br{\"a}nnstr{\"o}m, K.~Sou, and Y.~Sui, ``Radio
  resource management for {D2D-based V2V} communication,'' \emph{IEEE Trans.
  Veh. Technol.}, vol.~65, no.~8, pp. 6636--6650, Aug. 2016.

\bibitem{Sun2016cluster}
W.~Sun, D.~Yuan, E.~G. Str{\"o}m, and F.~Br{\"a}nnstr{\"o}m, ``Cluster-based
  radio resource management for {D2D}-supported safety-critical {V2X}
  communications,'' \emph{IEEE Trans. Wireless Commun.}, vol.~15, no.~4, pp.
  2756--2769, Apr. 2016.

\bibitem{Ashraf2018dynamic}
M.~I. Ashraf, C.-F. Liu, M.~Bennis, and W.~Saad, ``{Dynamic resource allocation
  for optimized latency and reliability in vehicular networks},'' \emph{IEEE
  Access}, vol.~6, pp. 63\,843--63\,858, Oct. 2018.

\bibitem{Liang2017resource}
L.~Liang, G.~Y. Li, and W.~Xu, ``{Resource allocation for D2D-enabled vehicular
  communications},'' \emph{IEEE Trans. Commun.}, vol.~65, no.~7, pp.
  3186--3197, Jul. 2017.

\bibitem{Liang2017spectrum}
L.~Liang, J.~Kim, S.~C. Jha, K.~Sivanesan, and G.~Y. Li, ``{Spectrum and power
  allocation for vehicular communications with delayed CSI feedback},''
  \emph{IEEE Wireless Comun. Lett.}, vol.~6, no.~4, pp. 458--461, Aug. 2017.

\bibitem{Liang2018graph}
L.~Liang, S.~Xie, G.~Y. Li, Z.~Ding, and X.~Yu, ``{Graph-based resource sharing
  in vehicular communication},'' \emph{IEEE Trans. Wireless Commun.}, vol.~17,
  no.~7, pp. 4579--4592, Jul. 2018.

\bibitem{Ye2018mlv2x}
H.~Ye, L.~Liang, G.~Y. Li, J.~Kim, L.~Lu, and M.~Wu, ``Machine learning for
  vehicular networks: Recent advances and application examples,'' \emph{IEEE
  Veh. Technol. Mag.}, vol.~13, no.~2, pp. 94--101, Jun. 2018.

\bibitem{Liang2018toward}
L.~Liang, H.~Ye, and G.~Y. Li, ``{Toward intelligent vehicular networks: A
  Machine Learning Framework},'' \emph{IEEE Internet Things J.}, vol.~6, no.~1,
  pp. 124--135, Feb. 2019.

\bibitem{salahuddin2016reinforcement}
M.~A. Salahuddin, A.~Al-Fuqaha, and M.~Guizani, ``Reinforcement learning for
  resource provisioning in the vehicular cloud,'' \emph{IEEE Wireless Commun.},
  vol.~23, no.~4, pp. 128--135, Jun. 2016.

\bibitem{zheng2016delay}
Q.~Zheng, K.~Zheng, H.~Zhang, and V.~C.~M. Leung, ``Delay-optimal virtualized
  radio resource scheduling in software-defined vehicular networks via
  stochastic learning,'' \emph{IEEE Trans. Veh. Technol.}, vol.~65, no.~10, pp.
  7857--7867, Oct. 2016.

\bibitem{He2017integrated}
Y.~He, N.~Zhao, and H.~Yin, ``{Integrated networking, caching and computing for
  connected vehicles: A deep reinforcement learning approach},'' \emph{IEEE
  Trans. Veh. Technol.}, vol.~67, no.~1, pp. 44 -- 55, Jan. 2018.

\bibitem{atallah2017reinforcement}
R.~Atallah, C.~Assi, and J.~Y. Yu, ``A reinforcement learning technique for
  optimizing downlink scheduling in an energy-limited vehicular network,''
  \emph{IEEE Trans. Veh. Technol.}, vol.~66, no.~6, pp. 4592--4601, Jun. 2017.

\bibitem{atallah2017deep}
R.~Atallah, C.~Assi, and M.~Khabbaz, ``Deep reinforcement learning-based
  scheduling for roadside communication networks,'' in \emph{Proc. IEEE WiOpt},
  May 2017, pp. 1--8.

\bibitem{li2017user}
Z.~Li, C.~Wang, and C.-J. Jiang, ``User association for load balancing in
  vehicular networks: An online reinforcement learning approach,'' \emph{IEEE
  Trans. Intell. Transp. Syst.}, vol.~18, no.~8, pp. 2217--2228, Aug. 2017.

\bibitem{xu2014fuzzy}
Y.~Xu, L.~Li, B.-H. Soong, and C.~Li, ``Fuzzy {Q}-learning based vertical
  handoff control for vehicular heterogeneous wireless network,'' in
  \emph{Proc. IEEE ICC}, Jun. 2014, pp. 5653--5658.

\bibitem{Omidshafiei2017deep}
S.~Omidshafiei, J.~Pazis, C.~Amato, J.~P. How, and J.~Vian, ``Deep
  decentralized multi-task multi-agent reinforcement learning under partial
  observability,'' in \emph{Proc. Int. Conf. Mach. Learning (ICML)}, 2017, pp.
  2681--2690.

\bibitem{Foerster2017stabilising}
J.~Foerster, N.~Nardelli, G.~Farquhar, T.~Afouras, P.~H. S.~T. 1, P.~Kohli, and
  S.~Whiteson, ``{Stabilising experience replay for deep multi-agent
  reinforcement learning},'' in \emph{Proc. Int. Conf. Mach. Learning (ICML)},
  2017, pp. 1146--1155.

\bibitem{Nasir2018deep}
Y.~S. Nasir and D.~Guo, ``Deep reinforcement learning for distributed dynamic
  power allocation in wireless networks,'' \emph{arXiv preprint
  arXiv:1808.00490}, 2018.

\bibitem{tan1993multi}
M.~Tan, ``Multi-agent reinforcement learning: Independent vs. cooperative
  agents,'' in \emph{Proc. Int. Conf. Mach. Learning (ICML)}, 1993, pp.
  330--337.

\bibitem{tesauro2004extending}
G.~Tesauro, ``{Extending Q-learning to general adaptive multi-agent systems},''
  in \emph{Proc. Advances Neural Inf. Process. Syst. (NIPS)}, 2004, pp.
  871--878.

\bibitem{watkins1992q}
C.~J. Watkins and P.~Dayan, ``Q-learning,'' \emph{Machine Learning}, vol.~8,
  no. 3-4, pp. 279--292, May 1992.

\bibitem{3GPPsimulation}
R1-165704, \emph{{WF on SLS evaluation assumptions for eV2X}}, 3GPP TSG RAN WG1
  Meeting \#85, May 2016.

\bibitem{WINNER}
\emph{{WINNER II Channel Models}}, IST-4-027756 WINNER II D1.1.2 V1.2, Sep.
  2007. [Online]. Available:
  \url{http://projects.celtic-initiative.org/winner+/WINNER2-Deliverables/D1.1%
.2v1.1.pdf}.

\bibitem{ruder2016overview}
S.~Ruder, ``An overview of gradient descent optimization algorithms,''
  \emph{arXiv preprint arXiv:1609.04747}, 2016.

\end{thebibliography}

\end{document}